\documentclass[12pt,a4paper]{article}
\usepackage{epsfig}
\begin{document}
\begin{titlepage}
\begin{center}
{\hbox to\hsize{\hfill KEK-TH-880}}

\vspace{4\baselineskip}

\textbf{\Large 
 Lepton Flavor Violating Processes and Muon $g-2$ \\
\bigskip
in Minimal Supersymmetric SO(10) Model}
\bigskip
\bigskip
\vspace{2\baselineskip}

\textbf{Takeshi Fukuyama%
\footnote{E-Mail: fukuyama@se.ritsumei.ac.jp}
and Tatsuru Kikuchi%
\footnote{E-Mail: rp009979@se.ritsumei.ac.jp }
} \\ 
\bigskip
\textit{\small 
Department of Physics, Ritsumeikan University, Kusatsu, 
Shiga 525-8577, Japan 
}
\vspace{2\baselineskip}

\textbf{Nobuchika Okada%
\footnote{E-Mail: okadan@post.kek.jp} }\\
\bigskip
\textit{\small
Theory Group, KEK, Oho 1-1, Tsukuba, Ibaraki 305-0801, Japan}

\vspace{3\baselineskip}

\textbf{Abstract}\\
\end{center}
\noindent
In the recently proposed minimal supersymmetric SO(10) model, 
 the neutrino Dirac Yukawa coupling matrix, 
 together with all the other fermion mass matrices, 
 is completely determined 
 once free parameters in the model are appropriately fixed 
 so as to accommodate the recent neutrino oscillation data. 
Using this unambiguous neutrino Dirac Yukawa couplings, 
 we calculate the lepton flavor violating (LFV) processes 
 and the muon $g-2$ 
 assuming the minimal supergravity scenario. 
The resultant rates of the LFV processes 
 are found to be large enough 
 to well exceed the proposed future experimental bound, 
 while the magnitude of the muon $g-2$ can be 
 within the recent result by Brookhaven E821 experiment. 
Furthermore, we find that there exists a parameter region 
 which can simultaneously realize 
 the neutralino cold dark matter abundance
 consistent with the recent WMAP data. 
\end{titlepage}

\setcounter{footnote}{0}
\newpage
%
%
\section{Introduction} 
The problems of the observed solar neutrino deficit 
 and the atmospheric neutrino anomaly can be naturally solved 
 through the interpretation of the neutrino oscillations 
 among the different flavor neutrinos. 
Now the evidence of the neutrino masses 
 and flavor mixings are almost established \cite{review}, 
 and this is also the evidence of new physics
 beyond the standard model. 
Interestingly, the neutrino mass and mixing properties 
 have been revealed to be very different 
 from those of the other fermions, 
 namely, neutrino masses are very small 
 and the flavor mixing angles are very large. 
A new physics must explain them. 

The supersymmetric (SUSY) SO(10) grand unified model is 
 one of the attractive candidates for new physics. 
The experimental data support the unification 
 of the three gauge couplings 
 at the grand unified theory (GUT) scale 
 $M_G \sim 2 \times 10^{16}$ GeV 
 with the particle contents of 
 the minimal supersymmetric standard model (MSSM) 
 \cite{unification}. 
This model incorporates the right-handed neutrinos 
 as the  member of the {\bf 16} representation 
 together with other standard model fermions, 
 and provides the natural explanation of 
 the smallness of the neutrino masses  
 through the see-saw mechanism \cite{see-saw}. 

However, it is a non-trivial problem 
 whether a SO(10) model can simultaneously accommodate 
 all the observed quark-lepton mass matrix data, 
 since the quark and lepton mass matrices 
 are related with each other and severely constrained 
 by virtue of the quark-lepton unification into 
 the {\bf 16} representation. 
There are lots of works on this subject. 
Recently it has been shown \cite{Fukuyama-Okada} that 
 the so-called minimal supersymmetric SO(10) model 
 can simultaneously reproduce 
 all the observed quark-lepton mass matrix data 
 involving the neutrino oscillation data 
 by fixing free parameters in the model appropriately. 
In this model, 
 the neutrino Dirac Yukawa coupling matrix, 
 together with all the other fermion mass matrices, 
 is completely determined. 

The evidence of the neutrino flavor mixing implies that 
 the lepton flavor of each generation is not individually conserved. 
Therefore the lepton flavor violating (LFV) processes 
 in the charged-lepton sector such as 
 $\mu \rightarrow e \gamma$, $\tau \rightarrow \mu \gamma$ 
 are allowed. 
In simply extended models 
 so as to incorporate massive neutrinos into the standard model, 
 the rate of the LFV processes is accompanied 
 by a highly suppression factor, 
 the ratio of neutrino mass to the weak boson mass, 
 because of the GIM mechanism,  
 and is far out of the reach of the experimental detection. 
However, in supersymmetric models, the situation is quite different. 
In this case, soft SUSY breaking parameters can be new LFV sources, 
 and the rate of the LFV processes 
 are suppressed by only the scale 
 of the soft SUSY breaking parameters 
 which is assumed to be the electroweak scale. 
Thus the huge enhancement occurs compared to the previous case. 
In fact, the LFV processes can be one of the most important processes 
 as the low-energy SUSY search \cite{Hisano}. 

Any mechanism of the SUSY breaking mediation are normally considered 
 so as to induce the soft SUSY breaking parameters 
 being almost flavor blind and real 
 because of sever constraints on the SUSY 
 flavor and CP violating processes \cite{FCNC-CP}. 
Therefore the rate of the LFV processes is negligible 
 at the SUSY breaking mediation scale. 
However, in a model with the LFV interactions, 
 the sizable LFV sources at low energies may be induced 
 through the renormalization effects. 
As well-motivated candidates for such models, 
 the SUSY GUTs \cite{Barbieri} 
 and the models with see-saw mechanism \cite{Borzumati} 
 have been considered. 

The magnitude of the LFV rate depends on 
 the SUSY breaking mediation scenario and the LFV interactions. 
In this paper, we evaluate the rate of the LFV processes 
 in the minimal SUSY SO(10) model, 
 where the neutrino Dirac Yukawa couplings 
 are the primary LFV sources. 
The minimal supergravity (mSUGRA) scenario 
 is assumed as the SUSY breaking mediation mechanism. 
As mentioned above, 
 all the Yukawa couplings are determined 
 in the minimal SUSY SO(10) model, 
 we can give the definite predictions 
 for the rate of the LFV processes. 

Recent result of the muon $g-2$ from 
 Brookhaven E821 experiment \cite{E821} 
 reported the deviation from the standard model prediction. 
Although the uncertainty of the standard model prediction  
 has not been settled \cite{light-by-light}, 
 we would expect that the deviation lies in the range  
\begin{eqnarray}
 \delta a_{\mu} = a_{\mu}(E821)- a_{\mu}(SM)
 = (10-40) \times 10^{-10}   \; , 
 \label{g-2}
\end{eqnarray} 
where $a_{\mu}=(g_{\mu}-2)/2$. 
This may be the first evidence of the low-energy SUSY \cite{g-2SUSY}. 
Except for the flavor indeces, 
 both of the LFV processes such as $\mu \rightarrow e \gamma$ 
 and the muon $g-2$ originate from the same dipole-moment operators, 
 and there is a correlation between their magnitudes 
 \cite{Hisano-Tobe}. 
In the following, we calculate the SUSY contribution 
 to the muon $g-2$ with the same input parameters 
 as in the calculations of the LFV processes, 
 and show their correlations.

This paper is organized as follows: 
 in the next section, we give a brief review of 
 the minimal SUSY SO(10) model 
 developed in \cite{Fukuyama-Okada}, 
 and present the predicted neutrino Dirac Yukawa coupling matrix 
 which plays the central role in our results. 
In section 3,  using the neutrino Yukawa coupling matrix, 
 we calculate the rate of LFV processes and 
 the magnitude of the SUSY contribution to the muon $g-2$ 
 for various input parameters in mSUGRA scenario. 
The last section is devoted to summary.

\section{Minimal SO(10) model and its predictions} 

Here we give a brief review of the minimal SUSY SO(10) model%
\footnote{
There are other models which are also called  
``minimal SUSY SO(10) model'' 
in literatures \cite{minimal-II}. 
Such models include only {\bf 10} Higgs multiplet 
 and realize the Yukawa unification 
 between bottom quark and tau lepton. 
In our ``minimal'' model, a $\overline{\bf 126}$ Higgs multiplet 
 is additionally introduced, and 
 Yukawa unification is not necessarily satisfied. 
}
 recently reconsidered in Ref.~\cite{Fukuyama-Okada}, 
 and summarize its predictions. 
Even when we concentrate our discussion on the issue 
 how to reproduce the realistic fermion mass matrices in the SO(10) model, 
 there are lots of possibilities 
 for introduction of Higgs multiplets. 
The minimal supersymmetric SO(10) model is the one 
 where only one {\bf 10} and one $\overline{\bf 126}$ 
 Higgs multiplets have Yukawa couplings  (superpotential)  
 with {\bf 16} matter multiplets such as 
\begin{eqnarray}
 W_Y = Y_{10}^{ij} {\bf 16}_i H_{10} {\bf 16}_j 
           +Y_{126}^{ij} {\bf 16}_i H_{126} {\bf 16}_j \; , 
\label{Yukawa1}
\end{eqnarray} 
where ${\bf 16}_i$ is the matter multiplet of the $i$-th generation,  
 $H_{10}$ and $H_{126}$ are the Higgs multiplet 
 of {\bf 10} and $\overline{\bf 126} $ representations 
 under SO(10), respectively. 
Note that, by virtue of the gauge symmetry, 
 the Yukawa couplings, $Y_{10}$ and $Y_{126}$, 
 are, in general, complex symmetric $3 \times 3$ matrices. 

We assume some appropriate Higgs multiplets, 
 whose vacuum expectation values (VEVs) correctly 
 break the SO(10) GUT gauge symmetry into the standard model one 
 at the GUT scale, $M_G \sim  2 \times 10^{16}$ GeV. 
Suppose the Pati-Salam subgroup \cite{Pati-Salam}, 
 $G_{422}=SU(4)_c \times SU(2)_L \times SU(2)_R$, 
 at the intermediate breaking stage. 
Under this symmetry, 
 the above Higgs multiplets are decomposed as 
 ${\bf 10} \rightarrow 
 ({\bf 6},{\bf 1},{\bf 1}) + ({\bf 1},{\bf 2},{\bf 2}) $ 
 and 
 $\overline{\bf 126} \rightarrow 
 ({\bf 6}, {\bf 1}, {\bf 1} ) 
 + ( {\bf 10}, {\bf 3}, {\bf 1}) 
 + (\overline{\bf 10}, {\bf 1}, {\bf 3})  
 + ({\bf 15}, {\bf 2}, {\bf 2}) $, 
while ${\bf 16} \rightarrow ({\bf 4}, {\bf 2}, {\bf 1}) 
+ (\overline{\bf 4}, {\bf 1}, {\bf 2})$. 
Breaking down to the standard model gauge group, 
 $SU(4)_c \times SU(2)_R  \rightarrow SU(3)_c \times U(1)_Y$,   
 is accomplished by non-zero VEV  of 
 the $(\overline{\bf 10}, {\bf 1}, {\bf 3})$ Higgs multiplet. 
Note that the Majorana masses for the right-handed neutrinos 
 are also generated by this VEV 
 through the Yukawa coupling $Y_{126}$ 
 in Eq.~(\ref{Yukawa1}). 

After this symmetry breaking, 
 we find two pair of Higgs doublets 
 in the same representation as the pair in the MSSM. 
One pair comes from $({\bf 1},{\bf 2},{\bf 2}) \subset {\bf 10}$ 
 and the other comes from 
 $(\overline{\bf 15}, {\bf 2}, {\bf 2}) \subset \overline{\bf 126}$. 
Using these two pairs of the Higgs doublets, 
 the Yukawa couplings of Eq.~(\ref{Yukawa1}) are rewritten as 
\begin{eqnarray}
W_Y &=& 
 (u_R^c)_i  \left(
 Y_{10}^{ij}  H^u_{10} + Y_{126}^{ij}  H^u_{126}     
 \right) q_j 
+
 (d_R^c)_i  \left(
 Y_{10}^{ij}  H^d_{10} + Y_{126}^{ij}  H^d_{126}     
 \right) q_j  \nonumber \\ 
&+&
 (\nu_R^c)_i  \left(
 Y_{10}^{ij}  H^u_{10} - 3 Y_{126}^{ij} H^u_{126}   
 \right) \ell_j 
+
 (e_R^c)_i  \left(
 Y_{10}^{ij}  H^d_{10}  - 3 Y_{126}^{ij} H^d_{126}  
\right) \ell_j   \nonumber \\
&+&
 (\nu_R^c)_i  
 \left( Y_{126}^{ij} \; v_R \right) 
 (\nu_R^c)_j \;  , 
\label{Yukawa2}
\end{eqnarray} 
where $u_R$, $d_R$, $\nu_R$ and 
 $e_R$ are the right-handed $SU(2)_L$ 
 singlet quark and lepton superfields, $q$ and $\ell$ 
 are the left-handed $SU(2)_L$ doublet quark and lepton superfields, 
 $H_{10}^{u,d}$ and $H_{126}^{u,d}$ 
 are up-type and down-type Higgs doublet superfields 
 originated from $H_{10}$ and $H_{126}$, respectively, 
 and the last term is the Majorana mass term 
 of the right-handed neutrinos developed 
 by the VEV of the $(\overline{\bf 10}, {\bf 1}, {\bf 3})$ Higgs, $v_R$. 
The factor $-3$ in the lepton sector 
 is the Clebsch-Gordan coefficient. 

In order to preserve the successful gauge coupling unification, 
 suppose that one pair of Higgs doublets 
 given by a linear combination $H_{10}^{u,d}$ and $H_{126}^{u,d}$ 
 is light while the other pair is  heavy ($\geq M_G$).  
The light Higgs doublets are identified as 
 the MSSM Higgs doublets ($H_u$ and $H_d$) 
 and given by 
\begin{eqnarray} 
 H_u &=& \tilde{\alpha}_u  H_{10}^u  
      + \tilde{\beta}_u  H_{126}^u 
 \nonumber \\
 H_d &=& \tilde{\alpha}_d  H_{10}^d  
      + \tilde{\beta}_d  H_{126}^d  \; , 
 \label{mix}
\end{eqnarray} 
where $\tilde{\alpha}_{u,d}$ and $\tilde{\beta}_{u,d}$ 
 denote elements of the unitary matrix  
 which rotate the flavor basis in the original model 
 into the (SUSY) mass eigenstates. 
Omitting the heavy Higgs mass eigenstates, 
 the low energy superpotential is described 
 by only the light Higgs doublets $H_u$ and $H_d$ such that 
\begin{eqnarray}
W_Y &=& 
 (u_R^c) _i  
 \left( \alpha^u  Y_{10}^{ij} + 
        \beta^u   Y_{126}^{ij} \right)  H_u  q_j 
+
 (d_R^c)_i  
 \left( \alpha^d  Y_{10}^{ij} + 
        \beta^d   Y_{126}^{ij}  \right) H_d q_j  \nonumber \\ 
&+&
 (\nu_R^c)_i  
 \left( \alpha^u  Y_{10}^{ij} -3 
        \beta^u   Y_{126}^{ij} \right)  H_u  \ell_j 
+
 (e_R^c)_i  
 \left( \alpha^d  Y_{10}^{ij} -3 
        \beta^d   Y_{126}^{ij}  \right) H_d \ell_j \nonumber \\ 
 &+& 
 (\nu_R^c)_i  
   \left( Y_{126}^{ij} v_R \right)  (\nu_R^c)_j \; ,  
\label{Yukawa3}
\end{eqnarray} 
where the formulas of the inverse unitary transformation 
 of Eq.~(\ref{mix}), 
 $H_{10}^{u,d} = \alpha^{u,d} H_{u,d} + \cdots $ and 
 $H_{126}^{u,d} = \beta^{u,d} H_{u,d} + \cdots $, 
 have been used. 
Note that the elements of the unitary matrix, 
 $\alpha^{u,d}$ and $\beta^{u,d}$,   
 are in general complex parameters, 
 through which CP-violating phases are introduced 
 into the fermion mass matrices. 

Providing the Higgs VEVs, 
 $H_u = v \sin \beta$ and $H_d = v \cos \beta$ 
 with $v=174 \mbox{GeV}$, 
 the quark and lepton mass matrices can be read off as%
\footnote{
In general, the $SU(2)_L$ triplet Higgs in 
 $({\bf 10}, {\bf 3}, {\bf 1}) \subset \overline{\bf 126}$ 
 can obtain the VEV induced through the electroweak symmetry breaking
 and may play a crucial role of the light Majorana neutrino mass matrix. 
This possibility, so-called type II see-saw model, 
 has been discussed in detail in Ref.~\cite{Type-II}. } 
\begin{eqnarray}
  M_u &=& c_{10} M_{10} + c_{126} M_{126}   \nonumber \\
  M_d &=&     M_{10} +     M_{126}   \nonumber \\
  M_D &=& c_{10} M_{10} -3 c_{126} M_{126}   \nonumber \\
  M_e &=&     M_{10} -3     M_{126}   \nonumber \\
  M_R &=& c_R M_{126}  \; , 
 \label{massmatrix}
\end{eqnarray} 
where $M_u$, $M_d$, $M_D$, $M_e$, and $M_R$ 
 denote the up-type quark, down-type quark, 
 neutrino Dirac, charged-lepton, and 
 right-handed neutrino Majorana mass matrices, respectively. 
Note that all the quark and lepton mass matrices 
 are characterized by only two basic mass matrices, $M_{10}$ and $M_{126}$,   
 and three complex coefficients $c_{10}$, $c_{126}$ and $c_R$, 
 which are defined as 
 $M_{10}= Y_{10} \alpha^d v \cos\beta$, 
 $M_{126} = Y_{126} \beta^d v \cos\beta$, 
 $c_{10}= (\alpha^u/\alpha^d) \tan \beta$, 
 $ c_{126}= (\beta^u/\beta^d) \tan \beta $ and 
 $c_R = v_R/( \beta^d  v  \cos \beta)$), respectively.  
Except for $c_R$, 
  which is used to determine the overall neutrino mass scale, 
 this system has fourteen free parameters in total \cite{Matsuda-etal}, 
 and the strong predictability to the fermion mass matrices. 

Thirteen electroweak data 
 of six quark masses, three mixing angles and one phase 
 in the Cabibbo-Kobayashi-Maskawa (CKM) matrix, 
 and three charged-lepton masses 
 are extrapolated to the GUT scale 
 according to the renormalization group equations (RGEs) 
 with given $\tan \beta$, 
 and are used as inputs at the GUT scale.%
\footnote{
In our analysis, we have neglected SUSY threshold corrections 
 to the quark Yukawa coupling 
 whose importance in the case of large $\tan \beta$ 
 has been pointed out \cite{threshold}. 
In our following analysis with $\tan \beta =45$, 
 we find the corrections about 10\%, which would affect 
 our final results.  
However we can check that this effects can be compensated away 
 by slight changes of other input values, for example,
 mixing angles in CKM matrix within the experimental errors, 
 and our final results is found to be almost unchanged. 
} 
 Solving the GUT mass matrix relation among quarks and charged-leptons 
 obtained by Eq.~(\ref{massmatrix}), 
 we can describe the neutrino Dirac mass matrix $M_D$ 
 and $M_{126}$ as functions of only one free parameter $\sigma $, 
 the phase of the combination 
 $(3_{10}+c_{126})/(-c_{10}+c_{126})$. 
The light Majorana neutrino mass matrix, 
 $M_{\nu}=-M_D^T M_R^{-1} M_D = - c_R^{-1} M_D^T M_{126}^{-1} M_D$, 
 given through the see-saw mechanism 
 is extrapolated down to the electroweak scale 
 according to the RGE 
 for the effective dimension-five operator \cite{RGEdim5}, 
 and is compared to the neutrino oscillation data 
 for various $\sigma $. 
It has been shown \cite{Fukuyama-Okada} that 
 the appropriate value of $\sigma $ 
 can reproduce three neutrino flavor mixing angles 
 in the Maki-Nakagawa-Sakata (MNS) matrix 
 and the ratio of mass-squared differences 
 ( $\Delta m_{\odot}^2/ \Delta m_{\oplus}^2$) 
 consistent with the current neutrino oscillation data.  
Here $\Delta m_{\odot}^2$ and $\Delta m_{\oplus}^2$
 are the mass-squared differences 
 relevant for the solar and the atmospheric neutrino 
 oscillations, respectively.  
The parameter $c_R$ is used to determine 
 the overall scale of the right-handed Majorana neutrino mass,  
 and thus the scale of $\Delta m_\oplus^2$. 
Once the appropriate values of $\sigma$ and $c_R$ are chosen, 
 all the fermion mass matrix (or fermion Yukawa couplings) 
 are completely determined. 
Therefore the model has definite predictions 
 to physics related to the fermion Yukawa couplings. 

Although in Ref.~\cite{Fukuyama-Okada} 
 various cases with given $\tan \beta = 40-55$ have been analyzed, 
 we consider only the case $\tan \beta =45$ in the following. 
Our final result in the next section is almost insensitive 
 to $\tan \beta$ values in the above range. 
The predictions of the minimal SUSY SO(10) model 
 that we need in our calculation in the next section 
 are as follows \cite{Fukuyama-Okada}: 
 with $\sigma=3.198$ fixed, 
 the right-handed Majorana neutrino mass eigenvalues 
 are found to be (in GeV) 
 $M_{R_1}=1.64 \times 10^{11}$,  
 $M_{R_2}=2.50 \times 10^{12}$ and 
 $M_{R_3}=8.22 \times 10^{12}$, 
 where $c_R$ is fixed so that 
 $\Delta m_\oplus^2 = 2 \times 10^{-3} \mbox{eV}^2$. 
In the basis where both of the charged-lepton 
 and right-handed Majorana neutrino mass matrices 
 are diagonal with real and positive eigenvalues, 
 the neutrino Dirac Yukawa coupling matrix at the GUT scale 
 is found to be 
\begin{eqnarray}
 Y_{\nu} = 
\left( 
 \begin{array}{ccc}
-0.000135 - 0.00273 i & 0.00113  + 0.0136 i  & 0.0339   + 0.0580 i  \\ 
 0.00759  + 0.0119 i  & -0.0270   - 0.00419  i  & -0.272    - 0.175   i  \\ 
-0.0280   + 0.00397 i & 0.0635   - 0.0119 i  &  0.491  - 0.526 i 
 \end{array}   \right) \; .  
\label{Ynu}
\end{eqnarray}     
%

\section{LFV processes and muon $g-2$} 

Even though the soft SUSY breaking parameters are flavor blind 
 at the scale of the SUSY breaking mediation, 
 the LFV interactions in the model can induce 
 the LFV source at low energies 
 through the renormalization effects.
In the following analysis, 
 we assume the mSUGRA scenario \cite{mSUGRA} 
 as the SUSY breaking mediation mechanism. 
At the original scale of the SUSY breaking mediation, 
 we impose the following boundary conditions 
 characterized by only five parameters 
 (all of them are assumed to be real in this paper), 
 $m_0$, $M_{1/2}$, $A_0$, $B$ and $\mu$. 
Here, $m_0$ is the universal scalar mass, 
 $M_{1/2}$ is the universal gaugino mass, 
 and $A_0$ is the universal coefficient 
 of the trilinear couplings. 
The parameters in the Higgs potential, $B$ and $\mu$, 
 are determined at the electroweak scale 
 so that the Higgs doublets obtain 
 the correct electroweak symmetry breaking VEVs 
 through the radiative breaking scenario \cite{RBS}. 
The soft SUSY breaking parameters at low energies 
 are obtained through the RGE evolutions 
 with the boundary conditions. 

Although the SUSY breaking mediation scale 
 is normally taken to be the (reduced) Planck scale 
 or the string scale ($\sim 10^{18}$ GeV),
 in the following calculations 
 we impose the boundary conditions 
 at the GUT scale, and analyze the RGE evolutions 
 from the GUT scale to the electroweak scale. 
This ansatz is the same as the one in the so-called 
 constrained MSSM (CMSSM). 
Note that, even with only particle contents introduced 
 in the previous section, 
 the minimal SUSY SO(10) model 
 is not the asymptotic free gauge theory. 
When a complete set of the Higgs multiplets is taken into account, 
 the gauge coupling would quickly blow up just above the GUT scale. 
If this is the case, field theory description of the model 
 will be no longer meaningful above the GUT scale, and thus 
 it is reasonable to introduce the GUT scale cutoff into the model. 
If there exists new physics like M-theory \cite{M-Theory}, 
 where the string scale comes down to the GUT scale, 
 this treatment would be justified. 

The effective theory which we analyze below the GUT scale 
 is the MSSM with the right-handed neutrinos. 
The superpotential in the leptonic sector is given by 
\begin{eqnarray}
W_Y =  Y_{\nu}^{ij} (\nu_R^c)_i \ell_j H_u 
+ Y_e^{ij} (e_R^c)_i \ell_j H_d 
+ \frac{1}{2} M_{R_{ij}} (\nu_R^c)_i  (\nu_R^c)_j 
   + \mu H_d  H_u  \;  , 
 \label{Yukawa4}
\end{eqnarray} 
where the indeces $i$, $j$ run over three generations, 
 $H_u$ and $H_d$ denote the up-type and down-type MSSM Higgs doublets, 
 respectively, and  $M_{R_{ij}}$ is 
 the heavy right-handed Majorana neutrino mass matrix. 
We work in the basis where the charged-lepton Yukawa matrix 
 $Y_e$ and the mass matrix $M_{R_{ij}}$ 
 are real-positive and diagonal matrices:  
 $Y_e^{ij}=Y_{e_i} \delta_{ij}$ and  
 $M_{R_{ij}}=\mbox{diag} ( M_{R_1},  M_{R_2},  M_{R_3}) $. 
Thus the LFV is originated from the off-diagonal components 
 of the neutrino Dirac Yukawa coupling matrix $ Y_{\nu}$. 
The soft SUSY breaking terms in the leptonic sector 
 is described as 
\begin{eqnarray}
 -{\cal L}_{\mbox{soft}} &=& 
   \tilde{\ell}^{\dagger}_i 
   \left( m^2_{\tilde{\ell}} \right)_{ij}
   \tilde{\ell}_j 
 + \tilde{\nu}_{R i}^{\dagger} 
   \left( m^2_{ \tilde{\nu}} \right)_{ij}
    \tilde{\nu}_{R j} 
 + \tilde{e}_{R i}^{\dagger} 
   \left( m^2_{ \tilde{e}} \right)_{ij} 
\tilde{e}_{R j}    \nonumber  \\ 
&+& m_{H_u}^2 H_u^{\dagger} H_u + m_{H_d}^2 H_d^{\dagger} H_d  
+ \left(  B \mu H_d H_u 
 + \frac{1}{2} B_{\nu} M_{R_{ij}} 
 \tilde{\nu}_{R i}^{\dagger} \tilde{\nu}_{R j} 
+ h.c. \right)  \nonumber \\ 
&+& \left( 
  A_{\nu}^{ij} \tilde{\nu}_{R i}^{\dagger}  \tilde{\ell}_j H_u 
+ A_e^{ij} \tilde{e}_{R i}^{\dagger} \tilde{\ell}_j H_d  +h.c.  
 \right)  \nonumber \\ 
&+& \left( 
    \frac{1}{2} M_1 \tilde{B}  \tilde{B}  
 +  \frac{1}{2} M_2 \tilde{W}^a  \tilde{W}^a  
  + \frac{1}{2} M_3 \tilde{G}^a  \tilde{G}^a  + h.c. \right)  \; .
 \label{softterms} 
\end{eqnarray}

As discussed above, 
 we impose the universal boundary conditions at the GUT scale 
 such that 
\begin{eqnarray}
& & \left( m^2_{\tilde{\ell}} \right)_{ij} 
= \left( m^2_{ \tilde{\nu}} \right)_{ij}
=   \left( m^2_{ \tilde{e}} \right)_{ij} = m_0^2 \delta_{ij} \; , 
 \nonumber \\ 
& &m_{H_u}^2=m_{H_d}^2 = m_0^2  \; , 
  \nonumber \\ 
& & A_{\nu}^{ij} = A_0 Y_{\nu}^{ij}\; , \; \; 
  A_{e}^{ij} = A_0 Y_{e}^{ij} \; , 
  \nonumber \\ 
& & M_1=M_2=M_3= M_{1/2} \; ,  
\end{eqnarray}
and extrapolate the soft SUSY breaking parameters 
 to the electroweak scale according to their RGEs. 
The LFV sources in the soft SUSY breaking parameters 
 such as the off-diagonal components of 
 $\left( m^2_{\tilde{\ell}} \right)_{ij}$ and $A_{e}^{ij}$ 
 are induced by the LFV interactions 
 through the neutrino Dirac Yukawa couplings. 
For example, the LFV effect most directly emerges 
 in the left-handed slepton mass matrix 
 through the RGEs such as 
\begin{eqnarray}
\mu \frac{d}{d \mu} 
  \left( m^2_{\tilde{\ell}} \right)_{ij}
&=&  \mu \frac{d}{d \mu} 
  \left( m^2_{\tilde{\ell}} \right)_{ij} \Big|_{\mbox{MSSM}} 
 \nonumber \\
&+& \frac{1}{16 \pi^2} 
\left( m^2_{\tilde{\ell}} Y_{\nu}^{\dagger} Y_{\nu}
 + Y_{\nu}^{\dagger} Y_{\nu} m^2_{\tilde{\ell}} 
 + 2  Y_{\nu}^{\dagger} m^2_{\tilde{\nu}} Y_{\nu}
 + 2 m_{H_u}^2 Y_{\nu}^{\dagger} Y_{\nu} 
 + 2  A_{\nu}^{\dagger} A_{\nu} \right)_{ij}  \; ,
 \label{RGE} 
\end{eqnarray}
where the first term in the right hand side denotes 
 the normal MSSM term with no LFV. 
In the leading-logarithmic approximation, 
 the off-diagonal components ($i \neq j$)
 of the left-handed slepton mass matrix are estimated as 
\begin{eqnarray}
 \left(\Delta  m^2_{\tilde{\ell}} \right)_{ij}
 \sim - \frac{3 m_0^2 + A_0^2}{8 \pi^2} 
 \left( Y_{\nu}^{\dagger} L Y_{\nu} \right)_{ij} \; ,  
 \label{leading}
\end{eqnarray}
where the distinct thresholds of the right-handed 
 Majorana neutrinos are taken into account 
 by the matrix $ L = \log [M_G/M_{R_i}] \delta_{ij}$. 

The effective Lagrangian 
 relevant for the LFV processes ($\ell_i \rightarrow \ell_j \gamma$) 
 and the muon $g-2$ is described as 
\begin{eqnarray}
 {\cal L}_{\mbox{eff}}= 
 -  \frac{e}{2} m_{\ell_i} \overline{\ell}_j \sigma_{\mu \nu} F^{\mu \nu} 
 \left(A_L^{j i} P_L + A_R^{j i} P_R \right) \ell_i  \; , 
\end{eqnarray}
where $P_{R, L} = (1 \pm \gamma_5)/2 $ is  
 the chirality projection operator, 
 and  $A_{L,R}$ are the photon-penguin couplings of 1-loop diagrams 
 in which chargino-sneutrino and neutralino-charged slepton 
 are running. 
The explicit formulas of $A_{L,R}$ etc. used in our analysis 
 are summarized in \cite{Hisano-etal} \cite{Okada-etal}. 
The rate of the LFV decay of charged-leptons is given by 
\begin{eqnarray}
\Gamma (\ell_i \rightarrow \ell_j \gamma) 
= \frac{e^2}{16 \pi} m_{\ell_i}^5 
 \left( |A_L^{j i}|^2  +  |A_R^{j i}|^2  \right) \; , 
\end{eqnarray}
while the real diagonal components of $A_{L,R}$ 
 contribute to the anomalous magnetic moments of 
 the charged-leptons such as 
\begin{eqnarray}
 \delta a_{\ell_i}^{SUSY} = \frac{g_{\ell_i}-2}{2} 
  = -  m_{\ell_i}^2 
  \mbox{Re} \left[ A_L^{i i}  +  A_R^{i i}  \right]  \; . 
\end{eqnarray}
In order to clarify the parameter dependence 
 of the decay amplitude, 
 we give here an approximate formula of the LFV decay rate 
 \cite{Hisano-etal}, 
\begin{eqnarray}
\Gamma (\ell_i \rightarrow \ell_j \gamma) 
 \sim  \frac{e^2}{16 \pi} m_{\ell_i}^5 
 \times  \frac{\alpha_2}{16 \pi^2} 
 \frac{| \left(\Delta  m^2_{\tilde{\ell}} \right)_{ij}|^2}{M_S^8} 
 \tan^2 \beta \; , 
 \label{LFVrough}
\end{eqnarray}
where $M_S$ is the average slepton mass at the electroweak scale, 
 and $ \left(\Delta  m^2_{\tilde{\ell}} \right)_{ij}$ 
 is the slepton mass estimated in Eq.~(\ref{leading}). 
We can see that the neutrino Dirac Yukawa coupling matrix 
 plays the crucial role in calculations of the LFV processes. 
We use the neutrino Dirac Yukawa coupling matrix of Eq.~(\ref{Ynu})
 in our numerical calculations. 

Now we present our results 
 for the rate of the LFV processes and the muon $g-2$. 
In Fig.~1, the branching ratio of the muon LFV decay, 
 $\mbox{Br} (\mu \rightarrow e \gamma ) $, is plotted
 as a function of the universal scalar mass $m_0$ 
 for the various universal gaugino masses,  
 $M_{1/2}=400, 600, 800, 1000$ GeV, and $A_0=0$. 
The horizontal line denotes 
 the present experimental upper bound,
 $\mbox{Br} (\mu \rightarrow e \gamma ) \leq 1.2 \times 10^{-11}$ 
 \cite{PDG}. 
We always take $\mu >0$, which leads to 
 the SUSY contribution to the muon $g-2$ being positive. 
On the other hand, for various $m_0=400, 600, 800, 1000$ GeV 
 the branching ratio as a function of $M_{1/2}$ 
 is depicted in Fig.~2. 
The predicted branching ratios are large 
 so that a parameter space with small $m_0$ and $M_{1/2}$ 
 has already been excluded. 
Note that even for $m_0, M_{1/2} \sim 1$ TeV 
 the branching ratios well exceed 
 the future planed upper bound, 
 $\mbox{Br} (\mu \rightarrow e \gamma ) \leq 10^{-14}$ 
 \cite{Kuno-Okada}. 
The LFV process may be observed in the near future. 
The branching ratios for the process $\tau \rightarrow \mu \gamma$ 
 are plotted in Fig.~3 and 4 
 for the same input parameters in Fig.~1 and 2, respectively, 
 together with the current upper bound, 
 $\mbox{Br} (\tau \rightarrow \mu \gamma ) \leq 0.6 \times 10^{-6}$ 
  \cite{BELLE}. 
In this case, the resultant branching ratio is too small 
 to be accessible to the future sensitivity, 
 $\mbox{Br} (\tau \rightarrow \mu \gamma ) \leq 10^{-9}$ 
 \cite{BELLE}, for input parameters consistent 
 with the current bound on the muon LFV decay (see Fig.~1, 2). 
As can be understood from Eqs.~(\ref{leading}) and (\ref{LFVrough}), 
 the branching ratio depends on $A_0$. 
For fixed $m_0=600$ GeV and $M_{1/2}=800$ GeV, 
 the branching ratios of both 
 $\mu \rightarrow e \gamma$ and  $\tau \rightarrow \mu \gamma$ 
 are plotted in Fig.~5  as functions of $A_0$. 
The other LFV processes such as $\mu \rightarrow 3 e$ 
 and $\mu-e$ conversion rate in nuclei are also interesting, 
 since they are strongly constrained by experiments. 
These processes consist of 
 the 1-loop penguin-type and the box-type diagrams. 
In our parameter space, 
 contributions from penguin-type diagrams 
 involving the above $A_{L,R}$ dominates, 
 and we can find the approximation formulas, 
 $\mbox{Br}(\mu \rightarrow 3 e)/\mbox{Br}(\mu \rightarrow e \gamma) 
 \sim 7 \times 10^{-3}$ \cite{Hisano-etal} 
 and 
 $R(\mu \rightarrow e; \mbox{Ti} (\mbox{Al}) )
 /\mbox{Br}(\mu \rightarrow e \gamma) \sim 5 (3) \times 10^{-3}$ 
 (see, for example, Ref.~\cite{mue-conversion} 
 for detailed calculations) 
 for the $\mu-e$ conversion with the Ti (Al) target. 
These are close to the present experimental upper bounds, 
 $\mbox{Br}(\mu \rightarrow 3 e) \leq 1.0 \times 10^{-12}$ \cite{PDG}  
 and 
 $ R(\mu \rightarrow e; N) \leq 6.1 \times 10^{-13}$ \cite{SINDRUM}, 
 when $\mbox{Br}(\mu \rightarrow e \gamma)$ 
 is close to the present upper bound as seen in Fig.~1 and 2. 

Next let us discuss the muon $g-2$. 
Numerical results are depicted in Fig.~6 and 7 
 for the same input parameters in Fig.~1 and 2, respectively. 
Although these results are almost insensitive 
 to the neutrino Dirac Yukawa couplings and 
 the universal trilinear coupling $A_0$, 
 they are correlated with the LFV branching ratios 
 through the dependences on $m_0$ and $M_{1/2}$. 
Note that the input parameters providing 
 $\mbox{Br}(\mu \rightarrow e \gamma)$ 
 close to the present upper bound 
 predict the suitable magnitude for 
 the observed muon $g-2$ in Eq.~(\ref{g-2}). 

If the R-parity is conserved in SUSY models, 
 the lightest superpartner (LSP) is stable. 
The neutralino, if it is the LSP, 
 is the plausible candidate for the cold dark matter (CDM) 
 in the present universe. 
The recent Wilkinson Microwave Anisotropy Probe (WMAP) satellite data 
 \cite{WMAP}  
 provide estimations of various cosmological parameters 
 with greater accuracy. 
The current density of the universe is composed of 
 about 73\% of dark energy and 27\% of matter. 
Most of the matter density is in the form of 
 the CDM, and its density is estimated to be (in 2$\sigma$ range) 
\begin{eqnarray}
\Omega_{CDM} h^2 = 0.1126^{+0.0161}_{-0.0181} \; . 
 \label{WMAP} 
\end{eqnarray}
The parameter space of the CMSSM 
 which allows the neutralino relic density 
 suitable for the cold dark matter 
 has been recently re-examined 
 in the light of the WMAP data \cite{CDM}. 
It has been shown that the resultant parameter space 
 is dramatically reduced into the narrow stripe 
 due to the great accuracy of the WMAP data. 
It is interesting to combine this result 
 with our analysis of the LFV processes and the muon $g-2$. 
In the case relevant for our analysis, 
 $\tan \beta=45$, $\mu>0$ and $A_0=0$, 
 we can read off the approximate relation 
 between $m_0$ and $M_{1/2}$ 
 such as (see Figure 1 in the second paper of Ref.~\cite{CDM}) 
\begin{eqnarray}
m_0(\mbox{GeV}) = \frac{9}{28} M_{1/2}(\mbox{GeV}) + 150(\mbox{GeV}) \;,  
 \label{relation} 
\end{eqnarray} 
along which the neutralino CDM is realized.%
\footnote{
We can see that our resultant soft SUSY breaking mass spectrum 
 at the electroweak scale is almost the same as those in the CMSSM. 
Thus we use the CMSSM result of Eq.~(\ref{relation}) 
 as the neutralino CDM constraint. 
}  
$M_{1/2}$ parameter space is constrained within the range 
 $300 \mbox{GeV} \leq M_{1/2} \leq 1000 \mbox{GeV}$ 
 due to the experimental bound on the SUSY contribution 
 to the $ b \rightarrow s \gamma$ branching ratio 
 and the unwanted stau LSP parameter region. 
We show $\mbox{Br}(\mu \rightarrow e \gamma)$ and 
 the muon $g-2$ as functions of $M_{1/2}$ 
 in Fig.~8 and 9, respectively, 
 along the neutralino CDM condition of Eq.~(\ref{relation}). 
We find the parameter region, 
 $560 \mbox{GeV} \leq M_{1/2} \leq 800 \mbox{GeV}$, 
 being consistent with all the experimental data. 

Before closing this section, 
 we give a comment on the electric dipole moments (EDMs) 
 of the charged-leptons. 
If the diagonal components of $A_{L,R}$ have imaginary parts, 
 CP is violated and the EDMs of the charged-leptons are given by
\begin{eqnarray}
 d_{\ell_i}/e =  - m_{\ell_i} 
  \mbox{Im} \left[A_L^{ii}-A_R^{ii} \right] \; . 
\end{eqnarray}
These complex $A_{L,R}$ are induced through 
 the renormalization effects 
 in the same manner as for the LFV processes. 
Here the primary source of the CP-violation 
 is the CP-phases in the complex neutrino Dirac Yukawa coupling matrix 
 of Eq.~(\ref{Ynu}). 
For the same input parameter region analyzed above, 
 we obtain the results 
 of the electron and muon EDMs 
 such as (we find $d_e, d_{\mu} < 0$) 
 $|d_e| = 10^{-34}-10^{-33}$ [e cm] 
 and $ |d_{\mu}| = 10^{-31} - 10^{-30}$ [e cm], respectively, 
 which are far below the present experimental upper bounds \cite{PDG}, 
 $ d_e \leq 1.6 \times 10^{-27}$ [e cm] 
 and $ d_{\mu} \leq 3.7 \times 10^{-19}$ [e cm]. 
As an example of our results, 
 we present the electron EDM in Fig.~10 as a function of $M_{1/2}$ 
 along the cosmological constraint of Eq.~(\ref{relation}). 
If the measurement of the electron EDM could be improved 
 by six orders of magnitude, 
 $ d_e \leq 10^{-33}$ [e cm], as proposed in \cite{eEDM}, 
 there might be a chance to detect the electron EDM.

\section{Summary} 

The evidence of the neutrino masses and flavor mixings 
 implies the non-conservation of the lepton flavor symmetry 
 in each generations. 
Thus the LFV processes in the charged-lepton sector are allowed. 
In SUSY theory, there is the possibility 
 that the rate of the LFV processes will be accessible 
 to the current or future experimental bounds. 
The concrete information of the LFV interactions 
 is necessary to evaluate the rate of the LFV processes. 

It has recently been shown \cite{Fukuyama-Okada} that 
 the minimal SUSY SO(10) model 
 can simultaneously accommodate all the observed quark-lepton 
 mass matrix data involving the neutrino oscillation data
 with appropriately fixed free parameters. 
In this model, the neutrino Yukawa coupling matrix 
 are completely determined, 
 whose off-diagonal components are the primary source 
 of the lepton flavor violation 
 in the basis where the charged-lepton and the right-handed 
 neutrino mass matrices are real and diagonal. 
Using this Yukawa coupling matrix, 
 we have calculated the rate of the LFV processes 
 assuming the mSUGRA scenario. 
The branching ratio of the $\mu \rightarrow e \gamma$ process 
 is found to be large so that a parameter space 
 with small universal scalar and gaugino masses 
 has been already excluded 
 by the present experimental upper bound. 
Even for 1 TeV input parameters, 
 we found that the branching ratio well exceeds 
 the future planed experimental upper bound. 

We also have calculated the SUSY contribution to the muon $g-2$, 
 which is correlated with the rate of the LFV processes 
 through the input parameters in mSUGRA scenario. 
The resultant magnitude of the muon $g-2$ 
 is found to be within the range of 
 the recent result of Brookhaven E821 experiment 
 for the input parameters 
 providing the $\mu \rightarrow e \gamma$ branching ratio close
 to the current experimental bound. 

In CMSSM the parameter region realizing 
 the neutralino dark matter scenario 
 has been dramatically reduced by the recent WMAP data. 
We have performed the same analysis 
 for the LFV processes and the muon $g-2$ 
 by restricting the input parameters 
 within the cosmologically allowed region. 
We have found that there exists a parameter region 
 consistent with all the data. 
%

%
\newpage
\begin{figure}
\begin{center}
\epsfig{file=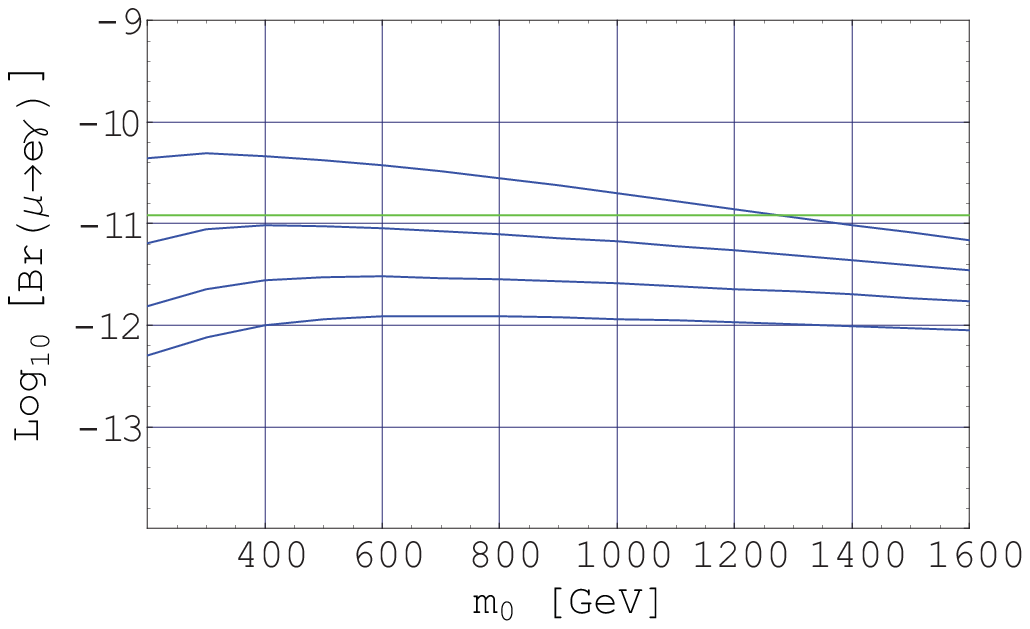, width=12cm}
\caption{
The branching ratio, 
$ \mbox{Log}_{10} \left[ \mbox{Br}(\mu \rightarrow e \gamma) \right]$, 
 as a function of $m_0$ (GeV) 
 for $M_{1/2} =400, 600, 800, 1000$ GeV (from top to bottom) 
 with $A_0=0$ and $\mu >0$. 
}
\end{center}
\end{figure}
\begin{figure}
\begin{center}
\epsfig{file=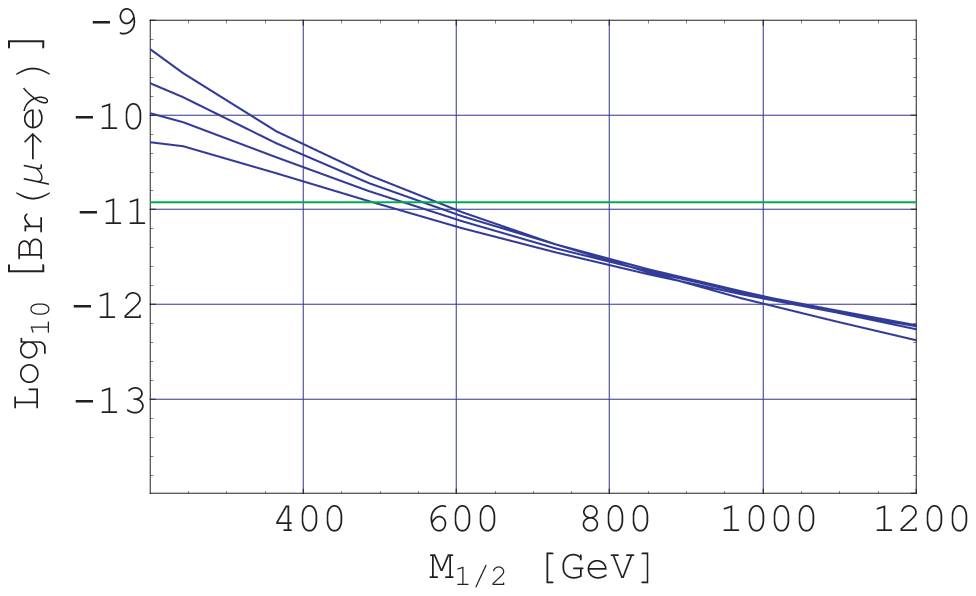, width=12cm}
\caption{
The branching ratio,  
$ \mbox{Log}_{10} 
\left[ \mbox{Br}(\mu \rightarrow e \gamma) \right] $,  
 as a function of $M_{1/2}$ (GeV) 
 for $m_0 =400, 600, 800, 1000$ GeV (from top to bottom) 
 with $A_0=0$ and $\mu >0$. 
}
\end{center}
\end{figure}
\begin{figure}
\begin{center}
\epsfig{file=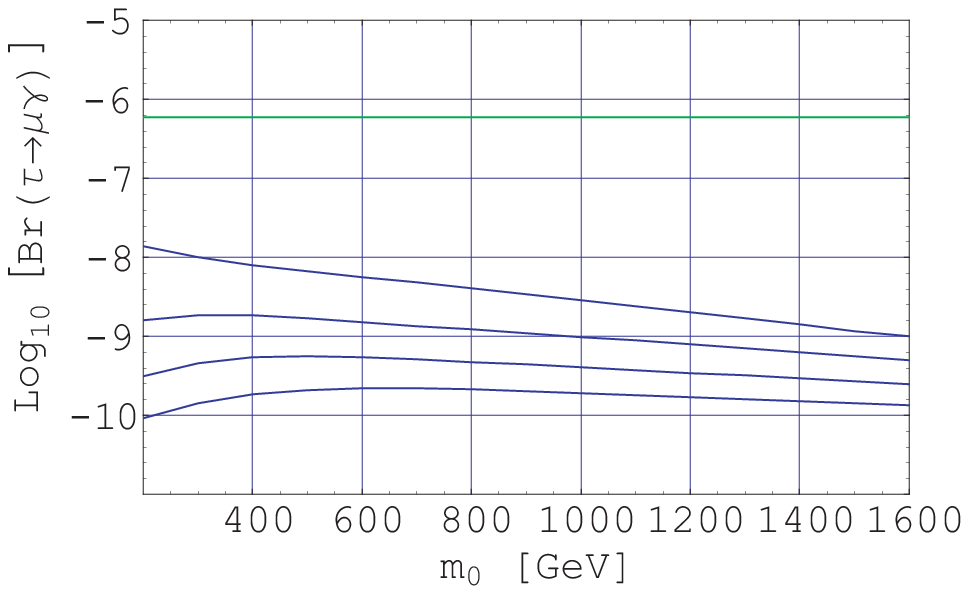, width=12cm}
\caption{
The branching ratio,  
$ \mbox{Log}_{10} 
\left[\mbox{Br}(\tau \rightarrow \mu \gamma)\right] $, 
 as a function of $m_0$ (GeV) 
 for $M_{1/2} =400, 600, 800, 1000$ GeV (from top to bottom) 
 with $A_0=0$ and $\mu >0$. 
}
\end{center}
\end{figure}
\begin{figure}
\begin{center}
\epsfig{file=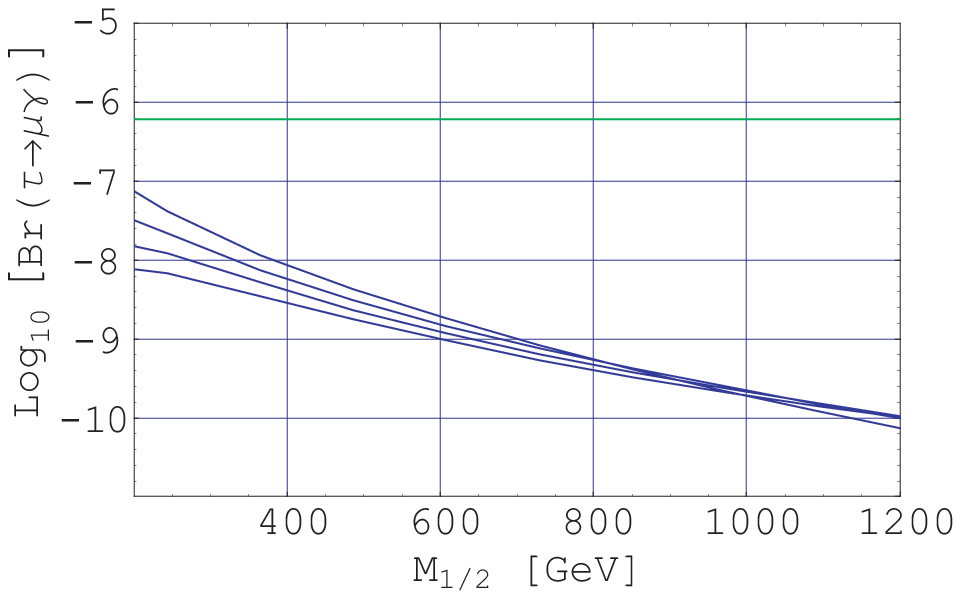, width=12cm}
\caption{
The branching ratio, 
$ \mbox{Log}_{10} 
 \left[ \mbox{Br}(\tau \rightarrow \mu \gamma) \right] $, 
 as a function of $M_{1/2}$ (GeV) 
 for $m_0 =400, 600, 800, 1000$ GeV (from top to bottom) 
 with $A_0=0$ and $\mu >0$. 
}
\end{center}
\end{figure}
\begin{figure}
\begin{center}
\epsfig{file=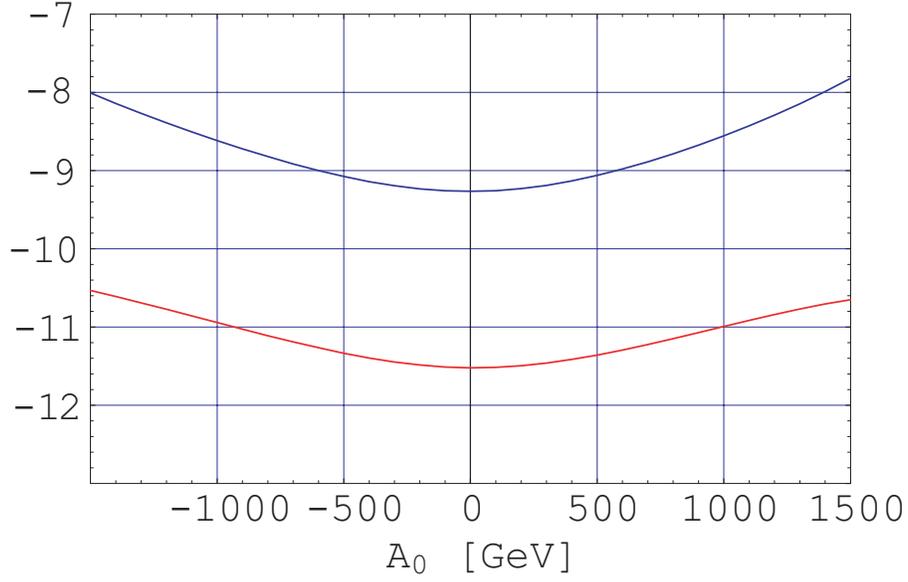, width=12cm}
\caption{
The branching ratios,  
$ \mbox{Log}_{10} \left[ \mbox{Br}(\tau \rightarrow \mu \gamma)
\right] $ (top)
and 
$ \mbox{Log}_{10} \left[ \mbox{Br}(\mu \rightarrow e \gamma)
\right] $ (bottom) 
 as functions of $A_0$ (GeV) 
 for $m_0 =600$ GeV and $M_{1/2}=800$ GeV. 
}
\end{center}
\end{figure}
\begin{figure}
\begin{center}
\epsfig{file=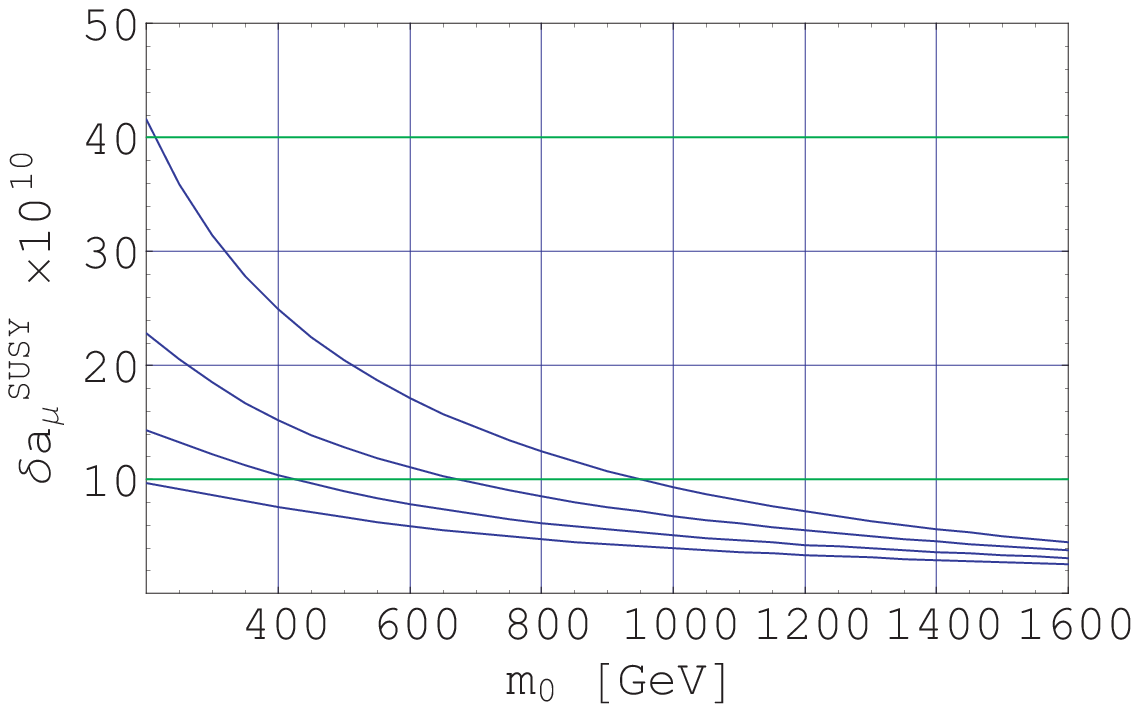, width=12cm}
\caption{
The SUSY contribution to the muon $g-2$ 
 in units of $10^{-10}$ 
 as a function of $m_0$ (GeV) 
 for $M_{1/2} =400, 600, 800, 1000$ GeV (from top to bottom)
 with $A_0=0$ and $\mu >0$. 
}
\end{center}
\end{figure}
\begin{figure}
\begin{center}
\epsfig{file=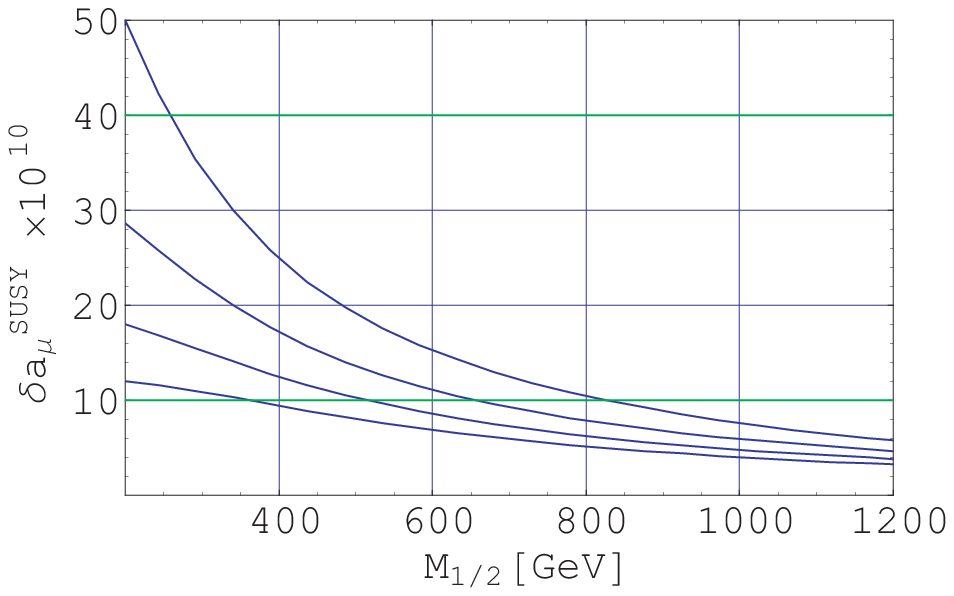, width=12cm}
\caption{
The SUSY contribution to the muon $g-2$ 
 in units of $10^{-10}$ 
 as a function of $M_{1/2}$ (GeV) 
 for $m_0 =400, 600, 800, 1000$ GeV 
 (from top to bottom) with $A_0=0$ and $\mu >0$. 
}
\end{center}
\end{figure}
\begin{figure}
\begin{center}
\epsfig{file=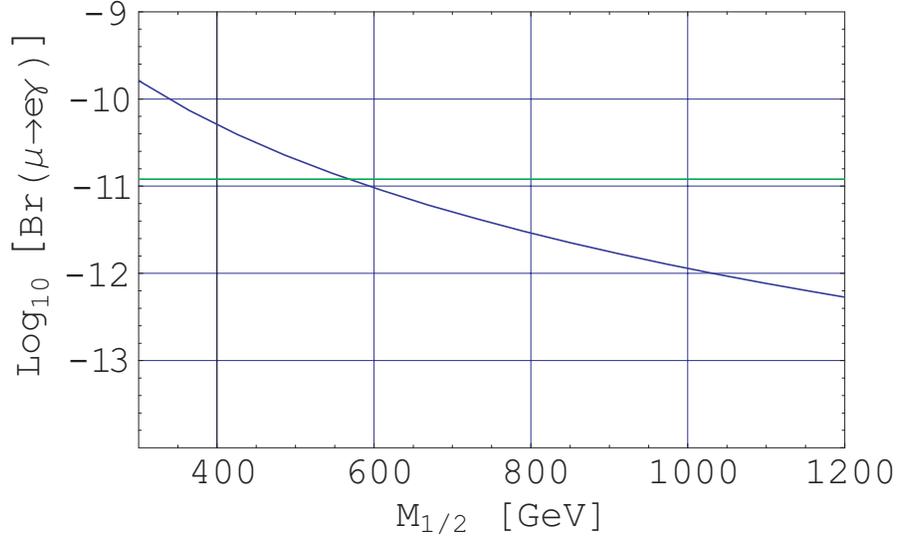, width=12cm}
\caption{
The branching ratio,  
$ \mbox{Log}_{10} 
\left[ \mbox{Br}(\mu \rightarrow e \gamma) \right]$,  
 as a function of $M_{1/2}$ (GeV) 
 along the cosmological constraint of Eq.~(\ref{relation}). 
}
\end{center}
\end{figure}
\begin{figure}
\begin{center}
\epsfig{file=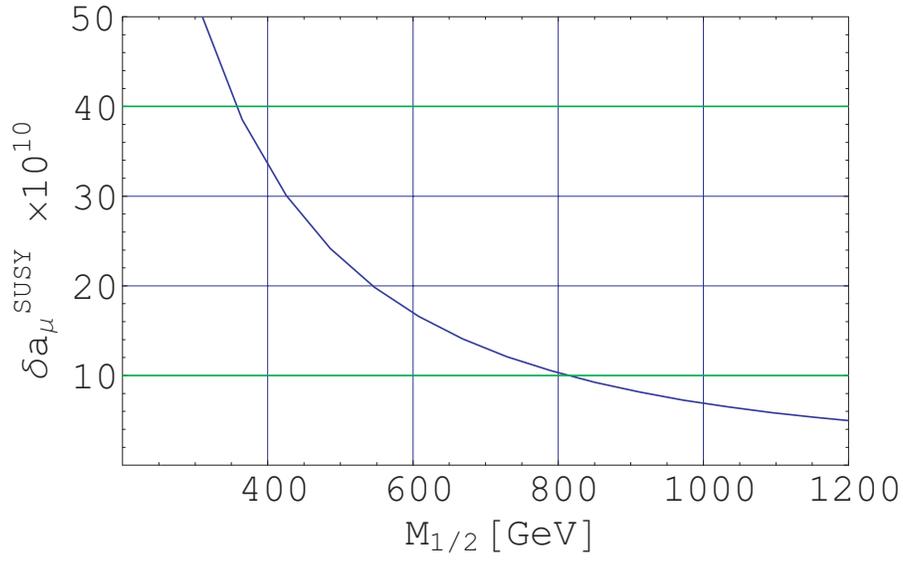, width=12cm}
\caption{
The SUSY contribution to the muon $g-2$ 
 in units of $10^{-10}$ 
 as a function of $M_{1/2}$ (GeV) 
 along the cosmological constraint of Eq.~(\ref{relation}). 
}
\end{center}
\end{figure}
\begin{figure}
\begin{center}
\epsfig{file=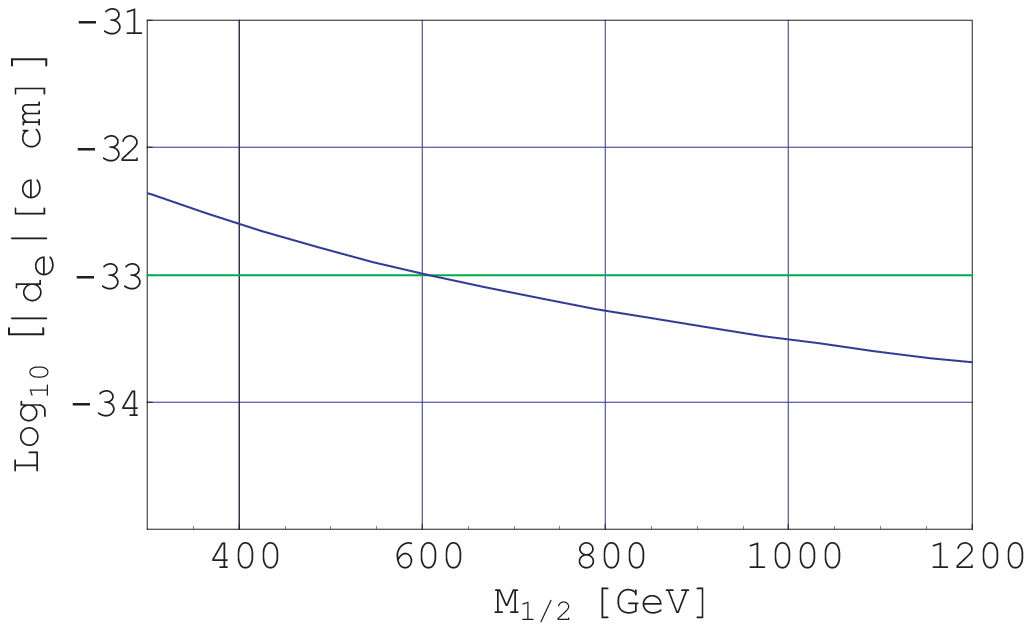, width=12cm}
\caption{
The electron EDM, 
 $ \mbox{Log}_{10} \left[ | d_e | [ \mbox{e cm}] \right] $, 
 as a function of $M_{1/2}$ (GeV) 
 along the cosmological constraint of Eq.~(\ref{relation}). 
}
\end{center}
\end{figure}

\begin{thebibliography}{99}
%
\bibitem{review}
For a recent review, see 
S.~Pakvasa and J.~W.~Valle, arXiv:hep-ph/0301061, 
and references therein.
%
\bibitem{unification}
C.~Giunti, C.~W.~Kim and U.~W.~Lee,
 Mod.\ Phys.\ Lett.\ A {\bf 6}, 1745 (1991); 
P.~Langacker and M.~x.~Luo,
 Phys.\ Rev.\ D {\bf 44}, 817 (1991); 
U.~Amaldi, W.~de Boer and H.~Furstenau,
 Phys.\ Lett.\ B {\bf 260}, 447 (1991).
%
\bibitem{see-saw} 
T. Yanagida, in Proceedings of the workshop 
 on the Unified Theory and Baryon Number in the Universe, 
 edited by O.Sawada and A.Sugamoto (KEK, Tsukuba, 1979);
M. Gell-Mann, P. Ramond, and R. Slansky, 
 in Supergravity, edited by D.Freedman and P.van Niewenhuizen 
 (north-Holland, Amsterdam 1979); 
R.~N.~Mohapatra and G.~Senjanovic,
Phys.\ Rev.\ Lett.\  {\bf 44}, 912 (1980).
%
\bibitem{Fukuyama-Okada} 
T.~Fukuyama and N.~Okada,
JHEP {\bf 0211}, 011 (2002). 
Also see arXiv:hep-ph/0205066 v2, 
 where some numerical errors in the version published in JHEP 
 have been corrected: 
 the mass eigenvalues $M_{R_i}$, and 
 $\langle m_{\nu} \rangle_{ee}$ and $\epsilon$ in Table 3. 
%
\bibitem{Hisano} 
For a recent review, see, for example, 
J.~Hisano, arXiv:hep-ph/0204100, 
and references therein. 
%
\bibitem{FCNC-CP}
For a general analysis see, for example, 
F.~Gabbiani, E.~Gabrielli, A.~Masiero and L.~Silvestrini,
 Nucl.\ Phys.\ B {\bf 477}, 321 (1996)
 [arXiv:hep-ph/9604387].
%
\bibitem{Barbieri}
For early works, see 
R.~Barbieri and L.~J.~Hall,
 Phys.\ Lett.\ B {\bf 338}, 212 (1994)
 [arXiv:hep-ph/9408406]; 
R.~Barbieri, L.~J.~Hall and A.~Strumia,
 Nucl.\ Phys.\ B {\bf 445}, 219 (1995)
 [arXiv:hep-ph/9501334]; 
Some of other relevant references are found in 
Ref.~\cite{Hisano}. 
%
\bibitem{Borzumati} 
For the early work, see 
F.~Borzumati and A.~Masiero,
Phys.\ Rev.\ Lett.\  {\bf 57}, 961 (1986); 
Some of other relevant references are found in 
Ref.~\cite{Hisano}.
%
\bibitem{E821} 
G.~W.~Bennett {\it et al.}  [Muon g-2 Collaboration],
 Phys.\ Rev.\ Lett.\  {\bf 89}, 101804 (2002)
 [Erratum-ibid.\  {\bf 89}, 129903 (2002)]
 [arXiv:hep-ex/0208001].
%
\bibitem{light-by-light}
M.~Knecht, A.~Nyffeler, M.~Perrottet and E.~De Rafael,
 Phys.\ Rev.\ Lett.\  {\bf 88}, 071802 (2002)
 [arXiv:hep-ph/0111059]; 
M.~Knecht and A.~Nyffeler,
 Phys.\ Rev.\ D {\bf 65}, 073034 (2002)
 [arXiv:hep-ph/0111058]; 
M.~Hayakawa and T.~Kinoshita,
 arXiv:hep-ph/0112102; 
J.~Bijnens, E.~Pallante and J.~Prades,
 Nucl.\ Phys.\ B {\bf 626}, 410 (2002)
 [arXiv:hep-ph/0112255]; 
I.~Blokland, A.~Czarnecki and K.~Melnikov,
 Phys.\ Rev.\ Lett.\  {\bf 88}, 071803 (2002)
 [arXiv:hep-ph/0112117].
%
\bibitem{g-2SUSY} 
There are numbers of papers on this subject. 
See, for example, the references 
 listed in Ref.\cite{Hisano}. 
%
\bibitem{Hisano-Tobe} 
J.~Hisano and K.~Tobe,
 Phys.\ Lett.\ B {\bf 510}, 197 (2001)
 [arXiv:hep-ph/0102315].
%
\bibitem{minimal-II} 
T.~Blazek, R.~Dermisek and S.~Raby,
Phys.\ Rev.\ Lett.\  {\bf 88}, 111804 (2002)
[arXiv:hep-ph/0107097]; 
T.~Blazek, R.~Dermisek and S.~Raby,
Phys.\ Rev.\ D {\bf 65}, 115004 (2002)
[arXiv:hep-ph/0201081].
%
\bibitem{Pati-Salam}
J.~C.~Pati and A.~Salam,
Phys.\ Rev.\ D {\bf 10}, 275 (1974).
%
\bibitem{Type-II}
B.~Bajc, G.~Senjanovic and F.~Vissani,
 Phys.\ Rev.\ Lett.\  {\bf 90}, 051802 (2003)
 [arXiv:hep-ph/0210207]; 
H.~S.~Goh, R.~N.~Mohapatra and S.~P.~Ng,
 arXiv:hep-ph/0303055.
%
\bibitem{Matsuda-etal}
K.~Matsuda, Y.~Koide and T.~Fukuyama,
 Phys.\ Rev.\ D {\bf 64}, 053015 (2001)
 [arXiv:hep-ph/0010026]; 
K.~Matsuda, Y.~Koide, T.~Fukuyama and H.~Nishiura,
 Phys.\ Rev.\ D {\bf 65}, 033008 (2002)
 [Erratum-ibid.\ D {\bf 65}, 079904 (2002)]
 [arXiv:hep-ph/0108202].
%
\bibitem{threshold}
L.~J.~Hall, R.~Rattazzi and U.~Sarid,
Phys.\ Rev.\ D {\bf 50}, 7048 (1994)
[arXiv:hep-ph/9306309]; 
M.~Carena, M.~Olechowski, S.~Pokorski and C.~E.~Wagner,
Nucl.\ Phys.\ B {\bf 426}, 269 (1994)
[arXiv:hep-ph/9402253]; 
R.~Hempfling,
Z.\ Phys.\ C {\bf 63}, 309 (1994)
[arXiv:hep-ph/9404226]; 
T.~Blazek, S.~Raby and S.~Pokorski,
Phys.\ Rev.\ D {\bf 52}, 4151 (1995)
[arXiv:hep-ph/9504364].

%
\bibitem{RGEdim5} 
P.~H.~Chankowski and Z.~Pluciennik,
 Phys.\ Lett.\ B {\bf 316}, 312 (1993)
 [arXiv:hep-ph/9306333]; 
K.~S.~Babu, C.~N.~Leung and J.~Pantaleone,
 Phys.\ Lett.\ B {\bf 319}, 191 (1993)
 [arXiv:hep-ph/9309223]; 
S.~Antusch, M.~Drees, J.~Kersten, M.~Lindner and M.~Ratz,
 Phys.\ Lett.\ B {\bf 519}, 238 (2001)
 [arXiv:hep-ph/0108005]; 
S.~Antusch, M.~Drees, J.~Kersten, M.~Lindner and M.~Ratz,
 Phys.\ Lett.\ B {\bf 525}, 130 (2002)
 [arXiv:hep-ph/0110366].
%
\bibitem{mSUGRA}
R.~Barbieri, S.~Ferrara and C.~A.~Savoy,
 Phys.\ Lett.\ B {\bf 119}, 343 (1982); 
A.~H.~Chamseddine, R.~Arnowitt and P.~Nath,
 Phys.\ Rev.\ Lett.\  {\bf 49}, 970 (1982); 
L.~J.~Hall, J.~Lykken and S.~Weinberg,
 Phys.\ Rev.\ D {\bf 27} (1983) 2359.
%
\bibitem{RBS}
K.~Inoue, A.~Kakuto, H.~Komatsu and S.~Takeshita,
 Prog. Theor. Phys. 68, 927 (1982);
L.~Iba\~nez and G.G.~Ross, Phys. Lett. 110B, 215 (1982);
L.~Alvarez-Gaume, M.~Claudson and M.B.~Wise,
 Nucl. Phys. B207, 96 (1982).
%
\bibitem{M-Theory}
E.~Witten,
 Nucl.\ Phys.\ B {\bf 471}, 135 (1996)
 [arXiv:hep-th/9602070]; 
P.~Horava and E.~Witten,
 Nucl.\ Phys.\ B {\bf 460}, 506 (1996)
 [arXiv:hep-th/9510209]; 
P.~Horava and E.~Witten,
 Nucl.\ Phys.\ B {\bf 475}, 94 (1996)
 [arXiv:hep-th/9603142].
%
\bibitem{Hisano-etal}
J.~Hisano, T.~Moroi, K.~Tobe and M.~Yamaguchi,
 Phys.\ Rev.\ D {\bf 53}, 2442 (1996)
 [arXiv:hep-ph/9510309].
%
\bibitem{Okada-etal}
Y.~Okada, K.~i.~Okumura and Y.~Shimizu,
 Phys.\ Rev.\ D {\bf 61}, 094001 (2000)
 [arXiv:hep-ph/9906446].
%
\bibitem{PDG}
K.~Hagiwara {\it et al.}  [Particle Data Group Collaboration],
 Phys.\ Rev.\ D {\bf 66}, 010001 (2002).
%
\bibitem{Kuno-Okada}
Y.~Kuno and Y.~Okada,
 Rev.\ Mod.\ Phys.\  {\bf 73}, 151 (2001)
 [arXiv:hep-ph/9909265].
%
\bibitem{BELLE}
K.~Inami, T.~Hokuue and T.~Ohshima  [BELLE Collaboration],
 [arXiv:hep-ex/0210036].
%
\bibitem{mue-conversion}
R.~Kitano, M.~Koike and Y.~Okada,
 Phys.\ Rev.\ D {\bf 66}, 096002 (2002)
 [arXiv:hep-ph/0203110].
%
\bibitem{SINDRUM}
P.~Wintz, in Proceedings of the First International Symposium 
 on Lepton and Baryon Number Violation, 
 edited by H. V. Klapdor-Kleingrothaus and I. V. Krivosheina 
 (Institute of Physics, Bristol, 1998), p. 534.
%
\bibitem{WMAP} 
C.~L.~Bennett {\it et al.},
 arXiv:astro-ph/0302207; 
D.~N.~Spergel {\it et al.},
 arXiv:astro-ph/0302209.
%
\bibitem{CDM}
J.~R.~Ellis, K.~A.~Olive, Y.~Santoso and V.~C.~Spanos,
 arXiv:hep-ph/0303043; 
A.~B.~Lahanas and D.~V.~Nanopoulos,
 arXiv:hep-ph/0303130.
%
\bibitem{eEDM}
S.~K.~Lamoreaux,
 arXiv:nucl-ex/0109014.
%
\end{thebibliography}
\end{document}